\documentclass{article}
\usepackage{spconf,amsmath,graphicx,multirow,booktabs,array,float,epstopdf,epsfig}

\title{Training Multi-Task Adversarial Network For Extracting Noise-Robust Speaker Embedding}
\name{Jianfeng Zhou$^1 $, Tao Jiang$^2$, Lin Li$^{*1}$, Qingyang Hong$^{*2}$, Zhe Wang$^3$, Bingyin Xia$^3$}

\address{$^1$College of Electronic Science and Technology, Xiamen University, China\\
	$^2$School of Information Science and Engineering, Xiamen University, China\\
	$^3$Media Coding Technology Lab, Huawei Media Technology Institute\\
	lilin@xmu.edu.cn, qyhong@xmu.edu.cn
}
\begin{document}
	%
	\maketitle
	\begin{abstract}
		Under noisy environments, to achieve the robust performance of speaker recognition is still a challenging task. Motivated by the promising performance of multi-task training in a variety of image processing tasks, we explore the potential of multi-task adversarial training for learning a noise-robust speaker embedding. In this paper, we present a novel framework that consists of three components: an encoder that extracts the noise-robust speaker embeddings; a classifier that classifies the speakers; a discriminator that discriminates the noise type of the speaker embeddings. Additionally , we propose a training strategy using the training accuracy as an indicator to stabilize the multi-class adversarial optimization process. We conduct our experiments on the English and Mandarin corpuses and the experimental results demonstrate that our proposed multi-task adversarial training method could greatly outperform the other methods without adversarial training in noisy environments. Furthermore, the experiments indicate that our method is also able to improve the speaker verification performance under the clean condition.
	\end{abstract}
	\section{Introduction}
	\label{sec:intro}
	
	The task of speaker verification is to verify the identity of speaker from a given speech utterance. In the past decade, the i-vector system has achieved significant success in modeling speaker identity and channel variability in the i-vector space \cite{001}, which maps variable-length utterances into fixed-length vectors. Then the fixed-length vectors will be fed to a back-end classifier such as probabilistic linear discriminant analysis (PLDA) \cite{002}.
	
	Recently, with the rise of deep learning \cite{003} in various machine learning applications, the works \cite{004,005,006} focused on using neural network to verify speakers have explored its potential capability in speaker recognition tasks. More recently, many studies \cite{007,008,009} have concentrated on extracting utterance-level representation, which is known as speaker embedding, using neural networks combined with a pooling layer. This utterance-level representation can be further processed by fully-connected layers.
	
	Since proposed by Goodfellow et al. \cite{010}, generative adversarial networks (GAN) have become the focus of many studies in recent years. Its great success in image processing has inspired people to consider whether it can also be applied into the field of speech processing. In the paper \cite{011}, Zhang et al. attempted to use conditional GAN to solve the impact of performance degradation caused by the variable-duration of utterances. Ding et al. \cite{012} proposed a multi-tasking GAN framework to extract the more distinctive speaker representation. And Yu et al. \cite{013} proposed to train an adversarial network for front-end denoising.
	
	In the field of speaker recognition, there is a large quantity of literature concerning the sharp degradation of performance in the noisy environments. A common way to improve the robustness of the system is to train the system using a dataset consisting of clean and noisy data \cite{014}. Speech enhancement is another way of denoising such as short-time spectral amplitude minimum mean square error (STSA-MMSE) \cite{015} and many DNN-based enhancement methods \cite{016,017,018}. Unlike previous works denoising in the front-end, we plan to use a multi-task adversarial framework to extract the noise-robust speaker representation directly.
	\begin{figure*}
		\centerline{\includegraphics[width=\textwidth]{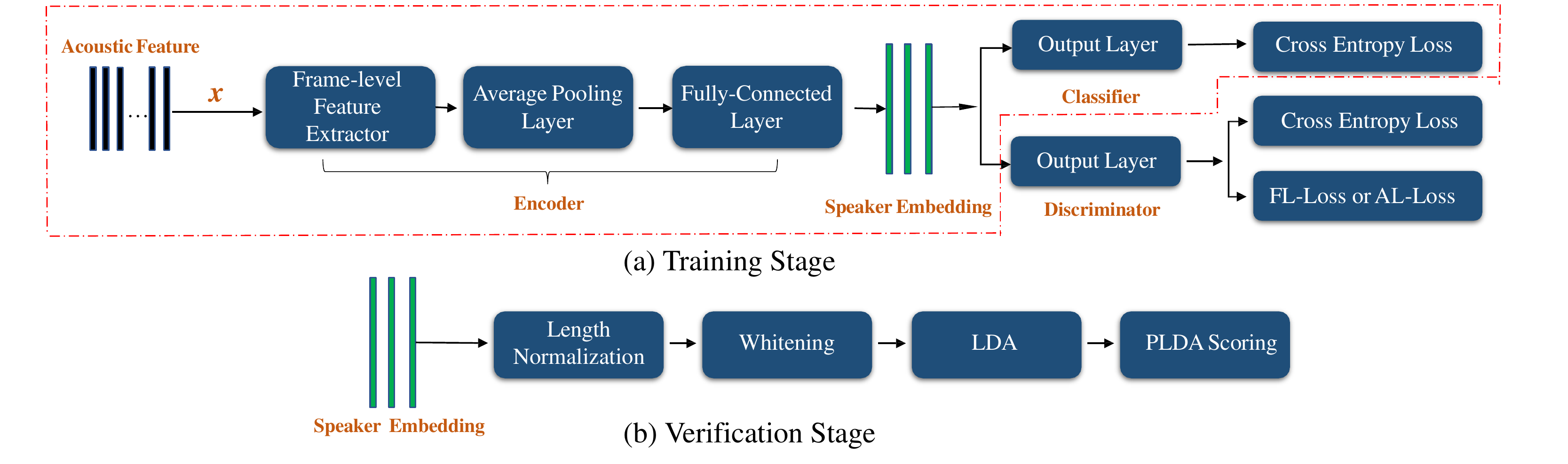}}
		\caption{The framework of our proposed multi-task adversarial network.}
		\label{fig:1}
	\end{figure*}
	
	In this paper, we borrow the adversarial training idea of GAN \cite{010} and use the multi-task adversarial network (MTAN) structure to extract a noise-robust speaker embedding. The entire framework consists of three parts: an encoder that extracts the noise-robust speaker embeddings; a classifier that classifies the speakers; a discriminator that discriminates the noise type of the speaker embeddings, which also plays the adversarial role combined with the encoder. In addition, we propose a new loss function, namely AL-Loss (anti-label loss), to realize the multi-class adversarial training. Furthermore, in order to balance the adversarial training process, a new training strategy has been presented by employing the training accuracy as an indicator to judge whether the adversarial training has reached a balance.
	\section{MULTI-TASK ADVERSARIAL NETWORK}
	\label{sec:format}
	\subsection{ CNN Based Embedding Learning}
	
	CNN-based neural network architecture has proved its superior performance in speaker verification tasks \cite{007,012}. In this work, we use the CNN-based architecture for speaker embedding learning which includes the encoder and classifier of the framework shown in the dotted line of Fig. \ref{fig:1} (a). The details of the architecture are as follow. Four one-dimensional convolutional layers with 1*1 filter, 1 stride and 256 channels followed by an average pooling layer which maps the frame-level feature to an utterance-level representation. Then, the speaker representation will be fed to the next two fully-connected layers with 256 and 1024 nodes in sequence. Finally, the output layer with $N_s$ (the number of speakers in training data) nodes will take the speaker embeddings as input. The output of last hidden layer is extracted as utterance-level speaker embedding. Besides, batch normalization and RELU activation function are applied to all layers except the output layer. And the verification back-ends are shown in Fig.\ref{fig:1} (b).
	
	\subsection{Multi-Task Adversarial Network}
	
	The entire architecture of MTAN is shown in Fig.\ref{fig:1} (a). And the implementation details of the encoder and classifier have been demonstrated in Section 2.1. As to the discriminator, it is just an output layer with $M$ (the number of noise types in training data) nodes. The arrows indicate the forward propagation direction.
	
	Given an input $ x \in {R^{t*m}} $ where $t$ and $m$ refer to the frame number and acoustic feature dimension of the utterance respectively, the encoder maps it to a speaker embedding $ E(x) \in {R^n} $, where $n$ is the dimension of latent embedding. Then the classifier and the discriminator try to predict the classes of $E(x)$. Since our goal is to encode speaker information while eliminating performance degradation caused by noise, the encoder should extract a latent representation that is more discriminative for speaker and robust for noise. In order to achieve this goal, we use the multi-task adversarial network to learn discriminative speaker feature and simultaneously improve its noise robustness. Specifically, we train the classifier cooperated with the encoder to extract discriminative speaker feature. Besides, we play a minimax game by training discriminator to maximize the probability of assigning the correct noise label to the embedding extracted from the encoder and simultaneously training the encoder to maximize the probability of assigning the wrong noise label to the embedding.
	\subsection{Loss Function}
	\label{ssec:subhead}
	In this work we consider cross entropy loss function and its two variants. For the cooperative training of the classifier and encoder, we directly minimize the cross entropy loss $l_s^C$ (the superscript $C$ means classifier). For multi-class adversarial training, the output of the discriminator will be fed to a cross entropy loss function $l_s^D$ (the superscript D means for discriminator) and its variants including FL-Loss (fixed label loss) proposed in \cite{013} and AL-Loss. The details of loss functions will be addressed in Section 2.3.1 and Section 2.3.2. Then a minimax game will be executed with the value function $l_{adv}$, which can be formulated as follow:
	\begin{equation}
	{\mathop {\max }\limits_E \mathop {\min }\limits_D  {l_{adv}} = \gamma {l_s^D} - \beta {l_{{\mathop{\textit {var}}} }} }
	\end{equation}
	where $\gamma$ and $\beta$ are scale parameters and $l_{\textit{var}}$ could be FL-Loss or AL-Loss. When training an adversarial network, rather than directly using the minimax loss, we split the optimization into two independent objectives, one for encoder and one for discriminator. Therefore, we train the encoder, discriminator and classifier by $\mathop {\min }\limits_E (l_s^C + \beta {l_{{\mathop{\textit{var}}} }})$, $\mathop {\min}\limits_D \gamma l_s^D$ and $\mathop {\min}\limits_C l_s^C$ respectively.
	\label{ssec:subhead}
	\subsubsection{FL-Loss}
	 Compared with the cross entropy loss function, FL-Loss uses the fixed label ``clean speech" \cite{013} for all inputs to train the encoder. It can be formulated as follow:
	\begin{equation}
	{l_{fl}} =  - \frac{1}{N}{\rm{ }}\sum\limits_{i = 1}^N {log} \frac{{{e^{W_{{y_c}}^T{y_i} + {b_{yc}}}}}}{{\sum\limits_{j = 1}^M {{{e^{W_j^T{y_i} + {b_j}}}}}}}
	\end{equation}
	where $N$ is the training batch size, $y_c$ is the label of clean speech and $y_i$ is the output of discriminator corresponding to $y_c$. Besides, $W$ and $b$ are the weights and biases of the output layer. By assigning all data to clean speech label, the embedding from noisy speech will be close to the embedding from clean speech, since the constraint of FL-Loss will regularize the encoder to learn a map function from noisy data distribution to clean data distribution.
	\subsubsection{AL-Loss}
	Inspired by the FL-Loss function, we propose the AL-Loss function combined with the cross entropy loss function for the multi-class adversarial task, which is formulated as follow:
	\begin{equation}
	{l_{al}} =  - \frac{1}{N}\sum\limits_{{\textit{i}} = 1}^N {\sum\limits_{j = 1,j \ne m_c}^M {\log \frac{{{e^{{W_j}^T{y_j}+{b_j} }}}}{{\sum\limits_{{\textit{k}} = 1}^M {{e^{W_k^T	{y_j} + {b_k}}}} }}} }
	\end{equation}
	where $m_c$ is the corresponding ground truth label of the $i$th sample. Unlike FL-Loss, we use the anti-label to calculate the loss value, where the anti-label means flipping the value of each bit in one hot vector of the ground truth label. $ \mathop {\min }\limits_E l_{anti} $ means that the encoder would be trained to assign the output of encoder to a wrong noise label equally, i.e., after adversarial training, the embedding extracted from encoder will be invariant to the clean and noisy speech.
	\section{EXPERIMENTS}
	\label{sec:pagestyle}
	\subsection{Dataset and Experimental Setting}
	To evaluate the effective performance of the proposed framework under the noisy environments, text-independent speaker verification (SV) experiments were conducted based on Aishell-1 \cite{019} (a Mandarin corpus) and Librispeech \cite{020} (an English corpus). The details of the two datasets are given as follows:
	\begin{itemize}
		\item \emph{Aishell-1}: We use the data of all three sets of Aishell-1 as the training data which contains about 141,600 utterances from 400 speakers and use another corpus named King-ASR-L-057\footnote{King-ASR-L-057: A Chinese Mandarin speech recognition database, which is available at  http://kingline.speechocean.com} as the test data which contains 6,167 recordings from 20 speakers.
		\item \emph{Librispeech}: In our experiments, we use the train-clean-500 part of Librispeech as training data which contains about 148,688 utterances from 1,166 speakers and the test-clean part as test data, which includes 2,020 recordings from 40 speakers.
	\end{itemize}

	We have made a noise corrupted version of the training data mentioned above by artificially adding different types of noise at different SNR levels. The original training data was divided into two parts with scale of 1:5, in which five out of six samples were added by the random noise. Speciﬁcally, the noisy utterances for training were made by adding one of the five noise types (white, babble, mensa, cafeteria, callcener)\footnote{white and babble were collected by Guoning Hu, and could be downloaded at http://web.cse.ohio-state.edu/pnl. Besides, cafeteria noise, callcener, and mensa were provided by HUAWEI TECHNOLOGIES CO., LTD.} randomly on the SNR levels of 10dB or 20dB. However, the noisy utterances for the speaker verification test were obtained by adding one of the five noise types on the SNR levels of 0dB, 5dB, 10dB, 15dB and 20dB respectively.
	
	All audios were converted to the features of 23-dimensional MFCC with a frame-length of 25 ms and the frame shift of 10 ms. Then, a frame-level energy-based voice activity detector (VAD) selection was conducted to the features. 
	
	Our implementation was based on the Tensorflow toolkit. In our experiments, Adam optimizer with a learning rate of 0.01 was used for the back propagation. We alternate between one step of optimizing the classifier and discriminator, and three steps of optimizing the encoder.
	
	\subsection{Training Stability}
	In this work, we use the training accuracy as an indicator to balance multi-class adversarial training. Specifically, we train the encoder to maximize the probability of assigning a speaker embedding to a wrong noise label, which means decreasing the training accuracy. However, we also train the discriminator to correctly assign the embedding to the ground truth label, which means increasing the training accuracy. So the accuracy could indicate the situation of adversarial training. The training accuracy keeping in high or low all means adversarial training doesn't get a balance. In addition, we set a lower threshold $\alpha$ and an upper threshold $\theta$. When the average of the training accuracy of the latest $K$ iterations is less than the lower threshold or higher than the upper threshold, we adjust the loss proportional factor of $\beta l_{var}$ and $\gamma l_s$ during the training. In our experiments, the encoder is trained better than discriminator, so we just set a lower threshold ($\alpha = 0.4$) to balance the adversarial training process.
	\subsection{Results and Comparisons}

In order to evaluate the performance of our proposed multi-task adversarial network, five systems were investigated: the CNN-based architecture trained using clean data (Baseline); the CNN-based architecture trained using the noise corrupted version of training data (MIX), which is a common method to improve the performance under noisy environments; MTAN trained using FL-Loss (FL); MTAN trained using AL-Loss (AL); the fusion system of FL and AL (Fusion). Specifically, the stabilization strategy proposed in this paper has been applied to both FL system and AL system. The equal error rate (EER) values of different methods are shown in Table 1 and Table 2. 
	\begin{table}
	\caption{EER(\%) of the SV system using four methods for different noise types and SNRs (dB) on Librispeech.}
	\scalebox{0.88}[0.86]{%
		\begin{tabular}{|c|cccccc|}
			\hline
			NOISE  & SNR & Baseline  & MIX  & FL  & AL  &Fusion\\ \hline
			Clean  &-    & 6.49  & 7.08 & 5.54  & 5.89    &\textbf{5.15}       \\ \hline
			\multirow{6}{*}{White}
			& 00    & 39.95 & 30.74 & 30.30 & 30.64  &\textbf{27.77}\\
			& 05    & 38.42 & 21.68 & 18.91 & 19.36 &\textbf{16.39}\\
			& 10    & 35.69 &15.25  & 12.23 & 13.07 &\textbf{10.35}\\
			& 15    & 29.50 & 12.23 & 9.90 & 10.35  &\textbf{8.71}\\
			& 20    & 24.26 & 10.89 & 8.86 & 9.46   &\textbf{7.77}\\ \cline{2-7}
			& mean  & 33.56 & 18.16 & 16.04 & 16.58 &\textbf{14.20}\\ \hline
			\multirow{6}{*}{Babble}
			& 00   & 30.74 & 20.05 & 20.00 & 18.71  &\textbf{17.72}\\
			& 05   & 25.05 & 12.72 & 11.09 & 11.19 &\textbf{10.30}\\
			& 10   & 19.46 & 10.00 & 8.07 & 8.32  &\textbf{7.77}\\
			& 15   & 14.41 & 8.91  & 7.53 & 7.72  &\textbf{6.93}\\
			& 20   & 11.09 & 8.07  & 6.49 & 6.54   &\textbf{6.09}\\ \cline{2-7}
			& mean & 20.10 & 11.95 & 10.64 & 10.50 &\textbf{9.76} \\ \hline
			\multirow{6}{*}{Cafeteria}
			& 00 & 32.52 & 19.80 & 20.30 & 18.91     &\textbf{17.18}\\
			& 05 & 26.73 & 14.36 & 12.03 & 12.72   &\textbf{10.74} \\
			& 10 & 21.24 & 10.99 & 9.26 & 9.41     &\textbf{8.27}\\
			& 15 & 16.14 & 8.91 & 7.48 & 7.62     &\textbf{6.83}\\
			& 20 & 12.03 & 8.37 & 6.24 & 6.93     &\textbf{6.09}\\ \cline{2-7}
			& mean & 21.73 & 12.49 & 11.06 & 11.12 &\textbf{9.82}\\ \hline
			\multirow{6}{*}{Callcener}
			& 00 & 28.81 & 15.79 & 14.85 & 14.31  &\textbf{13.27}\\
			& 05 & 23.12 & 10.00   & 9.21 & 10.00        &\textbf{8.76}\\
			& 10 & 17.28 & 8.71 & 7.48 & 7.33      &\textbf{6.63}\\
			& 15 & 12.67 & 7.97 & 6.24 & 6.63      &\textbf{5.89}\\
			& 20 & 9.90  & 7.72 & 6.49 & 6.29  	   &\textbf{5.89}\\	\cline{2-7}
			& mean & 18.36 & 10.04 & 8.85 & 8.91   &\textbf{8.09} \\ \hline
			\multirow{6}{*}{Mensa}
			& 00 & 35.89 & 21.14 & 20.05 & 20.30    &\textbf{18.56}\\
			& 05 & 31.14 & 14.16 & 11.68 & 13.12   &\textbf{10.64}\\
			& 10 & 25.10 & 9.75 & 9.11 & 9.31       &\textbf{8.07}\\
			& 15 & 19.21 & 8.71 & 7.23 & 7.67      &\textbf{6.68}\\
			& 20 & 14.11 & 7.87 & 6.14 & 6.68      &\textbf{6.04}\\ \cline{2-7}
			& mean & 25.09 & 12.33 & 10.84 & 11.42 &\textbf{10.00}\\ \hline
		\end{tabular}
	}
\end{table}

The results show that our proposed methods achieved the best performance across all of the SNR levels on Librispeech corpus and the lowest EERs across the majority of the SNR levels on Aishell-1 corpus. We can find that both FL system and AL system outperform the baseline and MIX system which indicates the adversarial training framework truly improves the performance of SV task under the noisy environments. Besides, we have conducted score-level fusion using the weights learned by linear regression algorithm to make full use of complementary information between FL system and AL system, which could further improve the discriminative ability of the system. In addition, the results on two corpuses in clean condition show that MTAN could outperform the Baseline system and MIX system even in the clean condition.
	
	\section{CONCLUSIONS}
\label{sec:majhead}
In this paper, we have explored the potential advantage of MTAN in extracting noise-robust speaker representation. The framework consists of three components: an encoder that extracts a noise-robust speaker embedding, a classifier and a discriminator that classifies the speaker and noise type of the speaker embedding respectively. Unlike the traditional multi-task learning where the encoder is trained to maximize the classification accuracy of the classifier and discriminator, MTAN is trained adversarially to the noise classification task, so that the embedding becomes speaker-discriminative and noise-robust. Experimental results on the Aishell-1 and Librispeech corpuses have shown that the proposed method could achieve dominant results in clean condition and the most noisy environments. In the future, we will conduct the experiments in lower SNR condition and other related applications.
	\begin{table}
	\caption{ EER(\%) of the SV system using four methods for different noise types and SNRs (dB) on Aishell-1.}
	\scalebox{0.88}[0.86]{%
		\begin{tabular}{|c|cccccc|c|}
			\hline
			NOISE  & SNR & Baseline  & MIX  & FL  & AL &Fusion  \\ \hline
			Clean  & -  & 7.33 & 10.39  & 4.63   & 4.64  & \textbf{3.82} \\ \hline
			\multirow{6}{*}{White}
			& 00     & 41.66    & \textbf{29.52}     & 36.01       & 34.60      &33.82     \\
			& 05     & 39.54    & \textbf{26.51}          & 30.83          & 27.42         & 27.03      \\
			& 10    & 36.14    & 24.28          & 24.23          &21.52          & \textbf{21.14} \\
			& 15    & 31.88    & 20.72          & 19.02          &17.75          & \textbf{16.02} \\
			& 20    & 26.30     & 17.90           & 14.86          &13.03          & \textbf{12.14} \\ \cline{2-7}
			& mean  & 35.10     & 23.79          & 24.99          &22.86          & \textbf{22.03} \\ \hline
			\multirow{6}{*}{Babble}
			& 00     & 28.48    & 24.49     & 25.73       & 25.55     &\textbf{22.93}     \\
			& 05     & 22.54    & 18.87          & 17.71          &17.56          & \textbf{15.44} \\
			& 10    & 17.76    & 15.59          & 12.72          &12.51          & \textbf{10.94} \\
			& 15    & 14.10     & 13.64          & 9.35           & 9.81          & \textbf{8.86} \\
			& 20    & 11.90     & 12.36          & 7.25           & 7.41          & \textbf{7.11} \\ \cline{2-7}
			& mean  & 18.96    & 16.99          & 14.55          & 14.57         & \textbf{13.02} \\ \hline
			\multirow{6}{*}{Cafeteria}
			& 00     & 29.24    & 24.75 & 25.15          & 25.64  & \textbf{22.58}        \\
			& 05     & 23.58    & 19.19          & 17.92          &17.27          & \textbf{15.41} \\
			& 10    & 18.60     & 15.86          & 12.54          &12.14          & \textbf{10.62} \\
			& 15    & 14.16    & 13.64          & 9.01           &8.92           & \textbf{8.04}  \\
			& 20    & 11.44    & 12.23          & 7.17           &6.88           & \textbf{6.62}  \\ \cline{2-7}
			& mean  & 19.40     & 17.13          & 14.36          &14.17          & \textbf{12.65} \\ \hline
			\multirow{6}{*}{Callcener} 
			& 00     & 27.24    & 22.71 & 23.48          & 22.95    & \textbf{20.47}      \\
			& 05     & 21.48    & 17.95          & 15.94          &15.88          & \textbf{13.61} \\
			& 10    & 16.72    & 14.87          & 11.75          &11.56         & \textbf{10.02} \\
			& 15    & 13.16    & 13.11          & 8.50            &8.42          & \textbf{7.83}  \\
			& 20    & 10.79    & 12.22          & 6.77           &6.68          & \textbf{6.49}  \\ \cline{2-7}
			& mean  & 17.88    & 16.17          & 13.29          &13.10          & \textbf{11.68}  \\ \hline
			\multirow{6}{*}{Mensa}     
			& 00     & 33.53    & 25.1  & 26.2           & 25.89     & \textbf{23.16}     \\
			& 05     & 27.84    & 20.07          & 18.76          &18.43         & \textbf{16.23} \\
			& 10    & 21.90     & 16.59          & 14.24          &13.69         & \textbf{12.07} \\
			& 15    & 16.90     & 14.26          & 10.55          &9.89          & \textbf{9.10}  \\
			& 20    & 13.61    & 12.61          & 8.12           &\textbf{7.56}          & 7.59  \\ \cline{2-7}
			& mean  & 22.76    & 17.73          & 15.57          &15.09         & \textbf{13.63} \\ \hline
		\end{tabular}
	}
\end{table}

\section{Acknowledgements}
	 This work was supported by the National Natural Science Foundation of China (Grant No.61876160).
	
	\bibliographystyle{IEEEbib}
	\bibliography{strings,refs}
\end{document}